\documentstyle[twoside,fleqn,npb,epsfig]{article}
\newcommand{\AmS}{{\protect\the\textfont2
  A\kern-.1667em\lower.5ex\hbox{M}\kern-.125emS}}

\title{Two-Loop Electroweak Corrections to $\Delta r$}

\author{A. Onishchenko\address{
State Research Center, Institute for High Energy Physics,\\
   Protvino, Moscow Region, 142284 Russia
   Mathematics and Computer Science Division}%
        ${}^{\rm b}$%
        \thanks{Supported
               by DFG-Forschergruppe  {\it ``Quantenfeldtheorie,
               Computeralgebra und Monte-Carlo-Simulation''}
               (contract FOR 264/2-1)}
        and
        O. Veretin\address{
      Institut f\"ur Theoretische Teilchenphysik,\\
       Universit\"at Karlsruhe, D-76128 Karlsruhe, Germany}%
        \thanks{Supported by BMBF under grant No 05HT9VKB0}%
     }

\begin{document}

\begin{abstract}
The two-loop electroweak bosonic correction to the muon lifetime
is computed using methods of asymptotic expansion. 
Combined with previous
calculations this completes the full two-loop correction
to $\Delta r$ in the Standard Model.
\end{abstract}

\maketitle

The Fermi constant $G_F$ plays an important role in the precision tests
of the Standard Model. Theoretically $G_F$ can be related to other
precision observables: the electroweak coupling constant $\alpha$
and the masses of
electroweak gauge bosons $M_Z$ and $M_W$. Other parameters enter
this expression through quantum corrections.
Usually one inverts this
relation in order to predict $M_W$ through $M_Z$ which is measured
much more accurate. This $M_Z-M_W$ interdependence can be then
confronted with experimental value $M_W^{\rm exp}$.
The current error (39 MeV) of $M_W^{\rm exp}$
will be drastically reduced at future colliders.
In fact, at LHC the experimental error can be reduced
to 15 MeV \cite{LHC} and at Linear Collider even down to 6 MeV \cite{TESLA}.
Therefore much efforts have been spent to reduce the error of
the theoretical prediction.

  The one-loop correction to $G_F$
is known since long ago \cite{DRoneloop}
along with the leading two-loop \cite{Largetop} corrections.
  Large two-loop
contributions from fermionic loops have been calculated in
\cite{Weiglein}.  The current prediction is affected by two
types of uncertainties. First, apart from the still unknown Higgs
boson mass, two input parameters introduce large errors. The current
knowledge of the top quark mass results in an error of about 30~MeV
\cite{Freitas:2002ja}, which should be reduced by LHC to 10~MeV and by
a linear collider even down to 1.2~MeV. The inaccuracy of the
knowledge of the running of the fine structure constant up to the
$M_Z$ scale, $\Delta \alpha(M_Z)$, introduces a further $6.5$~MeV
error. Second, several higher order corrections are unknown. In fact
the last unknown correction at the ${\cal O}(\alpha^2)$ order has been
calculated only recently in \cite{Czakon},\cite{Onishchenko:2002ve}
and \cite{vmeste}. 
This contribution comes from diagrams with no closed fermion loops.

Fermi constant is defined as the coupling constant
in the low energy four fermion effective Lagrangian
describing the decay of the muon
\begin{eqnarray}
\label{lagrangian}
{\cal L}_F = {\cal L}_{\rm QED}~~~~~~~~~~~~~~~~~~~~~~~~~~~~~~~~~~~
\\
\nonumber    
+~ \frac{G_F}{\sqrt{2}} \,
    \bigl[\bar{\nu}_\mu \gamma^\alpha (1-\gamma_5) \mu \bigr]
    \,   \bigl[  \bar{e} \gamma_\alpha (1-\gamma_5) \nu_e \bigr]
\end{eqnarray}
where $e$ and $\mu$ are electron and muon fields, $\nu_e$ and $\nu_\mu$
are the corresponding neutrinos and $G_F$ is the Fermi constant.
From the Lagrangian (\ref{lagrangian}) one gets
the following value for the muon lifetime
\begin{equation}
\frac{1}{\tau_\mu} =
\frac{G_F^2 m_\mu^5}{192\pi^3}
   \left( 1 - 8\frac{m_e^2}{m_\mu^2} \right) (1+\Delta q) \,,
\end{equation}
where the factor $\Delta q$ describes all the quantum corrections
in the low energy effective theory (i.e. QED corrections).
At one-loop order
these corrections have been computed a long time ago \cite{QEDoneloop}.
Recently also the two-loop result for $\Delta q$ has been obtained
\cite{QEDtwoloop}. Taking it into account,
the error of $G_F$ is nowadays dominated by the experimental error of
$\tau_\mu$ measurement\footnote{The present value is
$G_F = 1.16637(1)\times 10^{-5} \,\,\mbox{GeV}^{-2}$
and possible future experiments could reduce the error
by an order of magnitude.}.

  In order to relate $G_F$ to $M_W$ one can use
the matching condition between effective theory (\ref{lagrangian})
and the Standard Model, which requires that the value of $\tau_\mu$
does not depend on whether it is evaluated in the Fermi theory
or in the full Standard Model up to operators of higher dimensions,
i.e.
\begin{eqnarray}
\label{mateq1}
  A^{{\rm SM}} = \frac{G_F}{\sqrt2} \,
  \langle \mu| O_F | e\nu_\mu\bar{\nu_e} \rangle 
  + {\cal O}\left(\frac{m_\mu^4}{M_W^4}\right)  \,.
\end{eqnarray}
This equation however gets quantum corrections. In fact in the
l.h.s. of (\ref{mateq1}) there appear contributions coming
from both short and long distance. It the r.h.s. these contributions
should be separately absorbed in $G_F$ and the matrix element
$\langle O_F \rangle$ respectively. In order to make this in
the most easiest and elegant way the factorization theorem is used.

  In order to separate the short and long distance dynamics 
at the level of a single Feynmans diagram we use the large mass
expansion procedure \cite{asymptotic}. Namely, given the graf $F$, then
asymptotically
\begin{equation}
  \label{asym}
  F \sim \sum\limits_{H \subseteq F} S \cdot T(H)  \,,
\end{equation}
where the sum runs over all ``hard'' subgraphs $H$ of the diagram $F$;
$S$ is a ``soft'' subgraph obtained from $F$ by shrinking $H$ to a
point  and $T$ stands for the Taylor expansion (before integration!)
of $H$ with respect to all ``soft'' parameters.  The exact rules for
construction of hard subgraphs are discussed in details in
\cite{asymptotic}. 

  At the level of Feynman amplitude the situation is more involved,
since each diagram has its own subgraphs.
However it is possible to rearrange the action of asymptotic
expansion (\ref{asym}) in the sum of all diagrams such that it
exponentiates. Such rearrangement is based on the combinatorial
properties of $R$-operation and the special structure of 
the expansion (\ref{asym}). The rigorous prove of factorization
theorem can be found in \cite{Gorishnii}. 

\begin{figure}
\psfig{figure=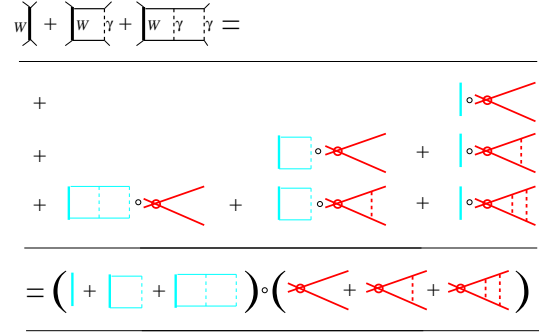,width=7cm}
\caption{\label{fact} Illustration of the factorization theorem.}
\end{figure}

In order to demonstrate the
idea we consider a simple example. In Fig. \ref{fact} the
contibution of ladder topologies to muon decay amplitude
$A^{\rm SM}$ is depicted.
One can notice however that this sum can be writen as a product
of two factors. One corresponds to short distance Wilson coefficient
function ($G_F$) and another to long distance matrix element.
All other topologies can be considered similary. As a result
of this procedure,
the evaluation of $G_F$ is reduced to computaion of
bubble diagrams only. 

  If we want to use the on-shell
scheme, the mass counterterms for $W$ and $Z$ bosons are needed.
They are given by the on-shell selfenergy diagrams.
This is the most difficult part of the calculation. However
in this case the asymptotic expansion can be applied.
In particular, the missing two-loop contributions to the mass
counterterms have been completed resently in \cite{bosons} 
(see also this proceeding).

  Therefore we have all ingredients to obtain the complete
two-loop electroweak correction to $G_F$. Explicit results
are given in \cite{Czakon,Onishchenko:2002ve,vmeste}.
In order to perform automatic calculation the simpbolic
program FORM \cite{FORM} and the diagram generator DIANA \cite{DIANA}
have been used. From the result of calculation,
in particular, it was found that the to the prediction of $M_W$
induced by the two-loop bosonic correction does not exceed 1~MeV
for the broad range of the Higgs boson mass from 100 to 1000~GeV.

In conclusion,
recent calculation of the two-loop bosonic corrections to $\Delta
r$ performed by two independent groups has been reviewed.
We concidered some details of the matching onto the Fermi theory.  
The framework for the evaluation of the Fermi constant $G_F$ based
on the low energy factorisation theorem has been constructed.
It allows one to compute $G_F$ as a Wilson coefficient in a simple manner.
This approach is general and is also applicable to other low energy 
quantities.



\begin{thebibliography}{99}

\bibitem{LHC}
{\it ATLAS: Detector and physics performance technical design report.
Vol. 2}, CERN-LHCC-99-15; ATLAS-TDR-15.

\bibitem{TESLA}
{\it TESLA: Technical design report. Part 3}, (eds. R. Heuer, D.J. Miller,
F. Richard and P.M. Zerwas), DESY-2001-011.

\bibitem{DRoneloop}
A. Sirlin, Phys.Rev. {\bf D22} (1980) 971.

\bibitem{Largetop}
G. Degrassi, P. Gambino and A. Vicini, Phys.Lett. {\bf B383} (1996) 219;\\
G. Degrassi, P. Gambino and A. Sirlin, Phys.Lett. {\bf B394} (1997) 188.

\bibitem{Weiglein}
A. Freitas et al., Phys.Lett. {\bf B495} (2000) 338;\\
A. Freitas et al., Nucl.Phys.Proc.Suppl. {\bf 89} (2000) 82.

\bibitem{Freitas:2002ja}
A.~Freitas, W.~Hollik, W.~Walter and G.~Weiglein,
Nucl.\ Phys.\ B {\bf 632} (2002) 189.

\bibitem{Czakon}
M. Awramik and M. Czakon, hep-ph/0208113.

\bibitem{Onishchenko:2002ve}
A.~Onishchenko and O.~Veretin,
arXiv:hep-ph/0209010.

\bibitem{vmeste}
M.~Awramik, M.~Czakon, A.~Onishchenko and O.~Veretin,
arXiv:hep-ph/0209084.

\bibitem{QEDoneloop}
S.M. Berman, Phys.Rev. {\bf 112} (1958) 267;\\
T. Kinoshita and A. Sirlin, Phys.Rev. {\bf 113} (1959) 1652.

\bibitem{QEDtwoloop}
T. van Ritbergen and R.G. Stuart, Phys.Rev.Lett. {\bf 82} (1999) 488;\\
T. van Ritbergen and R.G. Stuart, Nucl.Phys. {\bf B564} (2000) 343.



\bibitem{asymptotic}
F.V.~Tkachov, Preprint INR P-0332, Moscow (1983); P-0358, Moscow 1984;\\
K.G.~Chetyrkin,
Teor. Math. Phys. {\bf 75} (1988) 26; ibid {\bf 76} (1988) 207;
Preprint, MPI-PAE/PTh-13/91, Munich (1991);\\
~~V.A.~Smirnov,
Comm. Math. Phys.{\bf 134} (1990) 109;
{\it Renormalization and asymptotic expansions}
(Birkh\"auser, Basel, 1991);
{\it Applied asymptotic expansions in momenta and masses},
Berlin, Germany: Springer (2002), (Springer tracts in modern physics. 177).

\bibitem{Gorishnii}
S.G. Gorishnii, Nucl.Phys. {\bf B319} (1989) 633.

\bibitem{bosons}
F. Jegerlehner, M.Yu. Kalmykov and O. Veretin,
   Nucl.Phys. {\bf 641} (2002) 285; hep-ph/0105304 .

\bibitem{FORM}
J.A.M. Vermaseren, {\it Symbolic Manipulation with FORM},
Amsterdam, Computer Algebra, Netherland, 1991.

\bibitem{DIANA}
M. Tentyukov and J. Fleischer, Comput.Phys.Commun. {\bf 132} (2000) 124.

\end{thebibliography}
\end{document}